\documentclass[apj]{emulateapj}

\shorttitle{Fast Template Periodogram}
\shortauthors{Hoffman et al. 2021}

\usepackage[backref,breaklinks,colorlinks,citecolor=blue]{hyperref}
\usepackage[all]{hypcap}
\usepackage{amsmath}

\newcommand{\bigO}{\mathcal{O}}
\newcommand{\savg}[1]{\left<#1\right>}
\newcommand{\svar}{{\rm Var}}
\newcommand{\scov}{{\rm Cov}}
\newcommand{\Mshft}{\mathbf{M}_{\theta_2}}

\newcommand{\MMhat}{\widehat{MM}}
\newcommand{\YMhat}{\widehat{YM}}
\newcommand{\Mbar}{\overline{M}}

\newcommand{\YCt}{\widetilde{YC}}

\newcommand{\CCt}{\widetilde{CC}}
\newcommand{\CSt}{\widetilde{CS}}
\newcommand{\SSt}{\widetilde{SS}}

\newcommand{\ybar}{\bar{y}}
\newcommand{\Mbk}[1][k]{\Mbar^{(#1)}}
\newcommand{\Mt}{\mathbf{M}}
\newcommand{\Mk}[1][k]{M^{(k)}}
\newcommand{\Mtshft}{\Mt_{\theta_2}}
\newcommand{\YMh}[1][k]{\YMhat^{(#1)}}
\newcommand{\MMh}[1][k]{\MMhat^{(#1)}}
\newcommand{\yb}[1][k]{\ybar^{(#1)}}
\newcommand{\wk}[1][k]{w^{(#1)}}
\newcommand{\yhk}[1][k]{\hat{y}^{(#1)}}
\newcommand{\yk}[1][k]{y^{(#1)}}
\newcommand{\Mtshftk}[1][k]{\Mtshft^{(#1)}}
\newcommand{\Mtk}[1][k]{\Mt^{(#1)}}
\newcommand{\lk}[1][k]{\lambda^{(#1)}}
\newcommand{\Wtk}[1][k]{W^{(#1)}}

\newcommand{\eith}{\psi}

\begin{document}

\title{A Fast Template Periodogram for Detecting Non-Sinusoidal Fixed-Shape Signals in Irregularly Sampled Time Series}

\author{J. Hoffman\altaffilmark{1},
J. Vanderplas\altaffilmark{2},
J.~D. Hartman\altaffilmark{3},
G.~\'A Bakos\altaffilmark{3,}\altaffilmark{4}}
\email{johnh2o2@gmail.com}

\altaffiltext{1}{Xaxis, LLC, 3 World Trade Center, 175 Greenwich Street, 30th Floor, New York, NY 10007}
\altaffiltext{2}{eScience Institute, University of Washington, Seattle, WA 98195}
\altaffiltext{3}{Department of Astrophysical Sciences, Princeton University, Princeton NJ 08540}
\altaffiltext{4}{Institute for Advanced Study, 1 Einstein Drive, Princeton, NJ}

\begin{abstract}
    Astrophysical time series often contain periodic signals. The large and growing volume of time series data from
    photometric surveys demands computationally efficient methods for detecting and characterizing such signals.
    The most efficient algorithms available for this purpose are those that exploit the
    $\bigO(N\log N)$ scaling of the Fast Fourier Transform (FFT). However, these methods are not optimal
    for non-sinusoidal signal shapes. Template fits (or periodic matched filters) optimize
    sensitivity for \emph{a priori} known signal shapes but at a significant computational cost. Current
    implementations of template periodograms scale as $\bigO(N_f N_{\rm obs})$, where $N_f$ is the number
    of trial frequencies and $N_{\rm obs}$ is the number of lightcurve observations, and due to non-convexity, they do not
    guarantee the best fit at each trial frequency, which can lead to spurious results. In this work, we present a non-linear extension of the Lomb-Scargle
    periodogram to obtain a template-fitting algorithm that is both accurate (globally optimal solutions are
    obtained except in pathological cases) and computationally efficient (scaling as $\bigO(N_f\log N_f)$ for a given template).
    The non-linear optimization of the template fit at each frequency is recast as a polynomial
    zero-finding problem, where the coefficients of the polynomial can be computed efficiently with
    the non-equispaced fast Fourier transform. We show that our method, which uses truncated Fourier series to approximate templates,
    is an order of magnitude faster than existing algorithms for small problems ($N\lesssim 10$ observations) and
    2 orders of magnitude faster for long base-line time series with $N_{\rm obs} \gtrsim 10^4$ observations.
    An open-source implementation of the fast template periodogram is available at
    \href{https://www.github.com/PrincetonUniversity/FastTemplatePeriodogram}{github.com/PrincetonUniversity/FastTemplatePeriodogram}.
\end{abstract}

\section{Introduction}\label{sec:introduction}

Astronomical systems exhibit a wide range of time-dependent variability. By measuring and
characterizing this variability, astronomers are able to infer a variety of important
astrophysical properties about the underlying system.

Periodic signals in noisy astronomical timeseries can be detected with a number of techniques,
including Gaussian process regression \citep{Foreman-Mackey_etal_2017,Rasmussen+Williams_2006},
least-squares spectral analysis \citep{Lomb_1976,Scargle_1982,Barning_1963,Vanicek_1971}, and
information-theoretic methods \citep{Graham_etal_2013b,Huijse_etal_2012,Cincotta_etal_1995}. For an empirical
comparison of some of these techniques applied to several astronomical survey datasets, see
\cite{Graham_etal_2013}.

For stationary periodic signals, least-squares spectral analysis --- also known as
the Lomb-Scargle (LS) periodogram \citep{Lomb_1976,Scargle_1982,Barning_1963,Vanicek_1971} ---
is perhaps the most sensitive and computationally efficient method of detection. The LS periodogram
can be made to scale as $\bigO(N_f\log N_f)$, where $N_f$ is the number of trial frequencies, by
utilizing the non-equispaced fast Fourier transform \citep[][NFFT]{NFFT_KKD2009,NFFT_DR1993} to evaluate frequency-dependent
sums, or by ``extirpolating" irregularly spaced observations to a regular grid with Lagrange polynomials \citep{Press+Rybicki_1989}.

The LS periodogram fits the following model to a set of observations:

\begin{equation}
    \hat{y}_{\rm LS}(t|\theta, \omega) = \theta_0\cos{\omega t} + \theta_1\sin{\omega t}.
\end{equation}

where $\omega = 2\pi / P$ is the (angular) frequency of the underlying signal, and $\theta_0$ and
$\theta_1$ are the amplitudes of the signal. When the data $y_i$ is composed of a sinusoidal
component and a white noise component (i.e., when the measurement uncertainties are uncorrelated and
Gaussian), the LS periodogram provides a maximum likelihood estimate for the model parameters ($\omega, \theta_0,$ and
$\theta_1$).

The LS ``power'' $P_{LS}(\omega)$ has several definitions
in the literature \citep{Zechmeister+Kurster_2009}, but we adopt the following definition throughout the paper:

\begin{equation}
\label{eq:lsp}
P(\omega) = \frac{\chi^2_0 - \chi^2(\omega)}{\chi^2_0}
\end{equation}

Where $\chi^2_0$ is the weighted sum of squared residuals for a constant fit:

\begin{equation}
\chi^2_0 = \sum_i w_i (y_i - \bar{y})^2
\end{equation}

where $\bar{y} = \sum_i w_i y_i$ is the weighted mean of the observations, and
$w_i \propto \sigma_i^{-2}$ are the normalized weights for each observation ($\sum_i w_i = 1$),
and $\chi^2(\omega)$ is the weighted sum of squared residuals for the best-fit model $\hat{y}$ at
a given trial frequency $\omega = 2\pi / P$ where $P$ is the period:

\begin{equation}
\chi^2(\omega) = \min_{\theta} \sum_i w_i (y_i - \hat{y}(t_i| \omega, \theta))^2
\end{equation}

\subsection{Bayesian interpretation}

We note that, while this formalism captures many data and modeling scenarios, it is not completely general. For example, correlated
uncertainties are not handled here. A more general Bayesian treatment of periodic models in astronomical timeseries
is better handled by expressing a posterior over the model parameters.

Assuming a Gaussian likelihood for the observations $y_i \sim \mathcal{N}(\hat{y}( t_i | \omega, \theta), \sigma^2_i)$, and uniform
priors on both the frequency parameter $\omega$ and the non-frequency parameters $\theta$, the posterior is

\begin{align}
    p(\omega, \theta|X) &= \frac{p(X|\omega, \theta)p(\omega, \theta)}{p(X)}\\
                        &\propto p(\omega, \theta) \prod_{i=1}^{N_{\mathrm{obs}}}\mathcal{N}(\hat{y}(t_i | \omega, \theta), \sigma^2_i)
\end{align}

where we use $X = \{(t_i, y_i, \sigma_i) | 0 < i < N_{\mathrm{obs}}\}$ as shorthand for the lightcurve observations. The logarithm of the posterior is

\begin{align}
    \log p(\omega, \theta|X) &= -\frac{1}{2}\sum_{i=1}^{N_{\mathrm{obs}}}\left(\frac{y_i - \hat{y}(t_i | \omega, \theta)}{\sigma_i}\right)^2 + \mathrm{const.}\\
                             &= -\frac{W}{2}\sum_{i=1}^{N_{\mathrm{obs}}}w_i (y_i - \hat{y}(t_i | \omega, \theta))^2 + \mathrm{const.}
\end{align}

where $W= \sum_i \sigma_i^{-2}$. Thus, 

\begin{align}
    \chi^2(\omega) &= \min_{\theta} \sum_i w_i (y_i - \hat{y}(t_i|\omega, \theta))^2\\
                   &= \min_{\theta}\left(-\frac{2}{W}\log p(\omega, \theta|X) + \mathrm{const.}\right)\\
                   &= -\frac{2}{W}\max_\theta \log p(\omega, \theta|X) + \mathrm{const.}
\end{align}

and therefore, since $P(\omega) = 1 - \chi^2(\omega) / \chi_0^2$,

\begin{equation}
\label{eq:lsismap}
    P(\omega) = \frac{2}{W\chi_0^2} \max_\theta \log p(\omega, \theta|X) + \mathrm{const.}
\end{equation}

The Lomb-Scargle power is a linear transformation of the maximum of the log posterior over the non-frequency parameters, with the frequency parameter held fixed. Thus, choosing the frequency that maximizes the periodogram value corresponds to finding a MAP estimate of the frequency parameter.

A MAP interpretation is more general, and is applicable to scenarios not considered in this paper (e.g. correlated uncertainties, multi-dimensional timeseries, etc.), since all of these problems are amenable to MAP estimation of their model parameters. However, in order to maintain consistency with notation in the Lomb-Scargle literature \citep[e.g.][]{Zechmeister+Kurster_2009,Vanderplas_2018}, we keep our definition of the periodogram restricted as above and only consider one-dimensional timeseries with heteroscedastic but uncorrelated uncertainties.

\cite{Mortier_et_al_2015,Mortier_et_al_2015_code} explored a Bayesian interpretation of the periodogram. 
Their frequency periodogram definition marginalizes over non-frequency parameters in contrast to a 
(scaled) MAP value of the log likelihood discussed here. They use similar assumptions (uncorrelated, 
Normal measurement uncertainties) and use uniform priors on non-frequency parameters, but their 
periodogram power was based on marginal probabilities 
$P(\omega) = p(\omega|X) \propto \int_{\theta \in \Theta} p(\omega, \theta|X)p(\omega, \theta)$ 
where here the periodogram power is related to log-probabilities $\log p(\omega|X, \hat{\theta}(\omega))$ 
where $\omega$ is the angular frequency, $X$ is shorthand for observations, 
and $\hat{\theta}(\omega)$ are the non-frequency parameters that 
maximize $\log p(\omega, \theta|X)$ at a given trial frequency. 

Additionally, \cite{Mortier_et_al_2015} do not specifically consider the problem examined in this paper (periodic template fitting); they derive results for the classical periodogram model (sinusoidal model without constant offset) derived in
\cite{Lomb_1976} and \cite{Scargle_1982} as well as the extension with a constant offset derived in \cite{Zechmeister+Kurster_2009}. However, their general methodology (defining the periodogram power in terms of marginal probabilities), can be applied to our case and others.

As remarked in \citep{Vanderplas_2018}, Bayesian periodogram results should be taken with a grain of salt -- usually the underlying assumptions of the statistical model (uncorrelated, Normally-distributed measurement uncertainties, no systematics, etc.) are broken in many real-world astronomical timeseries..

\subsection{Extending Lomb-Scargle}

The LS periodogram has numerous extensions to account for, e.g., biased estimates of the mean brightness
\citep{Zechmeister+Kurster_2009}, non-sinusoidal signals \citep{Schwarzenberg-Czerny_1996,Bretthorst_2003,Palmer_2009,Baluev_2009,Baluev_2013a}, multiple periodicities \citep{Baluev_2013b,Baluev_2013c},
multi-band observations \citep{Vanderplas+Ivezic_2015}, and to mitigate overfitting of more
flexible models via regularization \citep{Vanderplas+Ivezic_2015}. \cite{Baluev_2008} provided the framework for improving false alarm statistics with extreme value theory, and this was generalized further in \cite{Baluev_2013a}. For a more detailed review
of the LS periodogram and its extensions, see \cite{Vanderplas_2018}.

The multi-harmonic LS periodogram \citep[][MHGLS]{Bretthorst+Chi-Cheng_1988,Schwarzenberg-Czerny_1996,Palmer_2009}, provides a more
flexible model by adding harmonic components to the fit:

\begin{equation}
    \hat{y}_{\rm MHLS}(t|\theta, \omega) = \theta_0 + \sum_{n=1}^H \theta_{2n}\cos{n\omega t} + \theta_{2n-1}\sin{n\omega t}.
\end{equation}

Additional harmonics are important for modeling signals that, while stationary and periodic,
are non-sinusoidal (e.g. RR Lyrae, eclipsing binaries, etc.).

However, the MHLS periodogram contains $2H+1$ free parameters while the original LS periodogram
contains only $2$ ($3$ if the mean brightness is considered a free parameter). Including higher order
harmonics adds model complexity, which can degrade the sensitivity of the MHLS periodogram to
sinusoidal or approximately sinusoidal signals.

Tikhonov regularization (or $L_2$ regularization), is one tool for mitigating overfitting
of the higher order harmonics. However, adding an $L_2$ regularization term to the Fourier amplitudes
adds bias to the model, and that bias should be compared to the value added by the
decrease in model variance \citep{Vanderplas+Ivezic_2015}.

\subsection{Computational scaling}

The LS periodogram naively scales as $\bigO(N_f N_{\rm obs})$ where $N_f$ is the number of trial
frequencies and $N_{\rm obs}$ is the number of observations. However, the limiting computations
of the LS periodogram involve sums of trigonometric functions over the observations. When
the observations are regularly sampled, the fast Fourier transform (FFT)  \citep{Cooley+Tukey_1965}
can evaluate such sums efficiently and the LS periodogram scales as $\bigO(N_f\log N_f)$.

When the data is not regularly sampled, as is the case for most astronomical time series,
the LS periodogram can be evaluated quickly in one of two popular ways.
The first, by \cite{Press+Rybicki_1989} involves``extirpolating'' irregularly sampled
data onto a regularly sampled mesh, and then performing FFTs to evaluate the necessary sums.
The second, as pointed out in \cite{Leroy_2012}, is to use the non-equispaced FFT \citep[][NFFT]{NFFT_KKD2009,NFFT_DR1993}
to evaluate the sums; this provides roughly an order of magnitude speedup over the \cite{Press+Rybicki_1989}
algorithm, and both algorithms scale as $\bigO(N_f\log N_f)$ \citep{Leroy_2012}.

There is a growing population of alternative methods for detecting
periodic signals in astrophysical data. Some of these methods can reliably
outperform the LS periodogram, especially for non-sinusoidal signal shapes
(see \cite{Graham_etal_2013} for a recent empirical review of period finding algorithms).
However, a key advantage the LS periodogram and its extensions is speed.
Virtually all other ``phase-folding" methods scale as $\bigO(N_{\rm obs}\times N_f)$, where $N_{\rm obs}$ is the number
of observations and $N_f$ is the number of trial frequencies, while the Lomb-Scargle
periodogram scales as $\bigO(N_f\log N_f)$. The virtual independence of Lomb-Scargle's computation
time with respect to the number of observations (assuming $N_f \gtrsim N_{\rm obs}$)
is especially valuable for lightcurves with $N_{\rm obs} \gg \log N_f \sim 50$.

Algorithmic efficiency will become increasingly important as the volume
of data produced by astronomical observatories continues to grow larger. The HATNet survey
\citep{HATNet}, for example, has already made $\bigO(10^4)$ observations of
$\bigO(10^6-10^7)$ stars. The Gaia telescope \citep{GAIA} is set to produce $\bigO(10-100)$
observations of $\bigO(10^9)$ stars. The Large Synoptic Survey Telescope (LSST; \cite{LSST})
will make $\bigO(10^2-10^3)$ observations of $\bigO(10^{10})$ stars during its operation starting in 2023.

\subsection{Template periodograms}

When the shape of a stationary periodic signal is known $\emph{a priori}$, then the number of degrees of freedom is the same as the original LS periodogram (with a floating mean component):

\begin{equation}
    \hat{y}(t|\theta, \omega) = \theta_0 + \theta_1 \mathbf{M}(\omega t - \theta_2),
\end{equation}

where $\mathbf{M}:[0, 2\pi)\rightarrow \mathcal{R}$ is a predefined periodic template.
We refer to the periodogram corresponding to this model as the ``template periodogram."

As is the case for the LS periodogram, the template periodogram
is equivalent to a maximum-likelihood estimate of the model parameters under the assumption that
measurement uncertainties are Gaussian and uncorrelated (i.e. white noise).

This paper develops new extensions of least-squares spectral analysis for arbitrary
signal shapes. For non-periodic signals this method is known as matched filter analysis,
and can be extended to search for periodic signals by, e.g., phase folding the data
at different trial periods.

An analysis by \cite{Sesar_etal_2016} found that template fitting significantly improved
period and amplitude estimation for RR Lyrae in Pan-STARRS DR1 photometry \citep{PanSTARRS}. Since the signal
shapes for RR Lyrae in various bandpasses are known \emph{a priori} (see \cite{Sesar_etal_2010}),
template fitting provides an optimal estimate of amplitude and period,
given that the object is indeed an RR Lyrae star well modeled by at least one of the templates.
Templates were especially crucial for Pan-STARRS data, since there are typically only
35 observations per source over 5 bands \citep{Hernitschek_etal_2016}, not enough to obtain
accurate amplitudes empirically by phase-folding. By including domain knowledge (i.e. knowledge of what RR Lyrae
lightcurves look like), template fitting allows for accurate inferences of amplitude even
for undersampled lightcurves.

However, the improved accuracy comes at substantial computational cost: the template fitting
procedure took 30 minutes per CPU per object, and \cite{Sesar_etal_2016} were forced to limit
the number of fitted lightcurves ($\lesssim 1000$) in order to keep the computational costs
to a reasonable level. Several cuts were made before the template fitting step to reduce the
more than 1 million Pan-STARRS DR1 objects to a small enough number, and each of these steps
removes a small portion of RR Lyrae from the sample. Though this number was reported by
\cite{Sesar_etal_2016} to be small ($\lesssim 2\%$), it may be possible to further improve
the completeness of the final sample by applying template fits to a larger number of objects,
which would require either more computational resources, more time, or, ideally, a more efficient
template fitting procedure.

The paper is organized as follows. Section \ref{sec:derivations} poses the problem of template
fitting in the language of least squares spectral analysis and derives the fast template
periodogram. Section \ref{sec:implementation} describes a freely available implementation
of the new template periodogram. Section \ref{sec:discussion} summarizes our results,
addresses caveats, and discusses possible avenues for improving the efficiency of the current
algorithm.

\section{Derivations}\label{sec:derivations}

We define a template $\mathbf{M}$

\begin{equation}
    \mathbf{M} : [0, 2\pi)\rightarrow\mathbb{R},
\end{equation}

\noindent as a mapping between the unit interval and the set of real numbers. We
restrict our discussion to sufficiently smooth templates such that
$\mathbf{M}$ can be adequately described by a truncated Fourier series

\begin{equation}
    \hat{\mathbf{M}}(\omega t|H) = \sum_{n=1}^H\left(c_n\cos{n\omega t} + s_n\sin{n\omega t}\right)
\end{equation}

\noindent for some finite $H > 0$.

That the $c_n$ and $s_n$ values are \emph{fixed} (i.e., they define
the template) is the crucial difference between the template periodogram and
the multi-harmonic Lomb-Scargle \citep{Palmer_2009,Bretthorst+Chi-Cheng_1988}, where $c_n$ and $s_n$
are \emph{free parameters}.

We now construct a periodogram for this template. The periodogram assumes
that an observed time series $S = \{(t_i, y_i, \sigma_i)\}_{i=1}^N$ can be modeled
by a scaled, transposed template that repeats with period $2\pi / \omega$, i.e.

\begin{equation}
y_i \approx \hat{y}(\omega t_i|\theta, \mathbf{M}) = \theta_1\mathbf{M}(\omega t_i - \theta_2) + \theta_3,
\end{equation}

\noindent where $\theta = (\theta_1, \theta_2, \theta_3)\in \mathbb{R}^3$ is a set of model parameters.

The optimal parameters are the location of a local minimum of the (weighted) sum of
squared residuals,

\begin{equation}
    \chi^2(\theta, S) \equiv \sum_i w_i (y_i - \hat{y}(\omega t_i|\theta) )^2,
\end{equation}

\noindent and thus the following condition must hold for all three model parameters at
the optimal solution $\theta=\theta_{\rm opt}$:

\begin{equation}\label{eq:chi2conds}
    \left.\frac{\partial\chi^2}{\partial\theta_j}\right|_{\theta=\theta_{\rm opt}} = 0~~\forall\theta_j\in\theta.
\end{equation}

Note that we have implicitly assumed $\chi^2(\theta,S)$ is a $C^1$ differentiable
function of $\theta$, which requires that $\mathbf{M}$ is a
$C^1$ differentiable function. Though this assumption could be violated if we
considered a more complete set of templates, (e.g. a box function), our restriction
to truncated Fourier series ensures $C^1$ differentiability.

Note that we also implicitly assume $\sigma_i > 0$ for all $i$ and we will later
assume that the variance of the observations $y_i$ is non-zero. If there are no
measurement errors, i.e. $\sigma_i = 0$ for all $i$, then uniform weights
(setting $\sigma_i = 1$) should be used. If the variance of the observations $y$
is zero, the periodogram (as defined in Equation \ref{eq:lsp}) is undefined for all frequencies. We do not
consider the case where $\sigma_i = 0$ for some observations $i$ and $\sigma_j > 0$ for
some observations $j$.

We can derive a system of equations for $\theta_{\rm opt}$ from the condition given
in Equation \ref{eq:chi2conds}. The explicit condition that must be met for each parameter $\theta_j$ is simplified below,
using

\begin{equation}
\hat{y}_i = \hat{y}(\omega t_i | \theta)
\end{equation}

\noindent and

\begin{equation}
\partial_j\hat{y}_i = \left.\frac{\partial \hat{y}(\omega t|\theta)}{\partial \theta_j}\right|_{t = t_i}
\end{equation}

\noindent for brevity:

\begin{equation}\label{eq:simpconds}
\begin{split}
0 &= \left.\frac{\partial\chi^2}{\partial\theta_j}\right|_{\theta=\theta_{\rm opt}}\\
  &= -2\sum_i w_i\left(y_i - \hat{y}_i \right)(\partial_j\hat{y})_i \\
\sum_i w_iy_i(\partial_j\hat{y})_i&= \sum_i w_i \hat{y}_i(\partial_j\hat{y})_i.
\end{split}
\end{equation}

The above is a general result that extends to all least squares periodograms.
To simplify derivations, we adopt the following notation:

\begin{eqnarray}
\savg{X} &\equiv& \sum_i w_i X_i\\
\savg{XY} &\equiv& \sum_i w_i X_iY_i\\
\scov(X, Y) &\equiv& \savg{XY} - \savg{X}\savg{Y}\\
\svar(X) &\equiv& \scov(X, X)
\end{eqnarray}

In addition, the transposed template $\Mshft = \mathbf{M}(\omega t_i - \theta_2)$ can be expressed
as

\begin{align}
\Mshft(\omega t) &= \sum_n c_n\cos n\left(\omega t - \theta_2 \right) \\
                &\qquad + s_n\sin{n\left(\omega t - \theta_2 \right)}\\
                &= \sum_n\left(c_n\cos{n\theta_2}-s_n\sin{n \theta_2}\right)\cos{n\omega t} + \\
                &\qquad \left(s_n\cos{n\theta_2} + c_n\sin{n \theta_2}\right)\sin{n\omega t} \\
                &= \sum_n\left(\alpha_n e^{in\theta_2} + \alpha_n^{*} e^{-in\theta_2}\right)\cos{n\omega t} + \\
                &\qquad \left(-i\left[\alpha_n e^{in\theta_2} - \alpha_n^{*} e^{-in\theta_2}\right]\right)\sin{n\omega t} \\
                &= \sum_n\left(\alpha_n \eith^n + \alpha_n^{*} \eith^{-n}\right)\cos{n\omega t} + \\
                &\qquad \left(-i\left[\alpha_n \eith^n - \alpha_n^{*} \eith^{-n}\right]\right)\sin{n\omega t} \\
                &= \sum_nA_n(\eith)\cos{n\omega t} + B_n(\eith)\sin{n\omega t}
\end{align}

where $\alpha_n = (c_n + is_n)/2$, and $\eith\equiv e^{i\theta_2}$ is a convenient change of variable.

We also define the following terms:

\begin{align}
\YMhat &= \savg{y\Mshft}\\
\MMhat &= \savg{\Mshft^2}\\
\Mbar &= \savg{\Mshft}\\
MM &= \svar(\Mshft) = \MMhat - \Mbar^2\\
YM &= \scov(\Mshft, y) = \YMhat - \bar{y}\Mbar
\end{align}

For a given phase shift $\theta_2$, the optimal amplitude and
offset are obtained from requiring the partial derivatives of the
sum of squared residuals, $\chi^2$, to be zero.

Namely, we obtain that
\begin{align}
0 = \frac{\partial\chi^2}{\partial\theta_1} &= 2\sum_iw_i(y_i - \hat{y}_i)\left(-\frac{\partial\hat{y}}{\partial\theta_1}\right)_i\\
    &= \sum_iw_i(y_i - \theta_1\Mshft - \theta_3)\Mshft\\
    &= \YMhat - \theta_1 \MMhat - \theta_3 \Mbar
\end{align}

and
\begin{align}
0 = \frac{\partial\chi^2}{\partial\theta_3} &= 2\sum_iw_i(y_i - \hat{y}_i)\left(-\frac{\partial\hat{y}}{\partial\theta_3}\right)_i\\
    &= \sum_iw_i(y_i - \theta_1\Mshft - \theta_3)\\
    &= \bar{y} - \theta_1 \Mbar - \theta_3
\end{align}

This system of equations can then be rewritten as

\begin{equation}
\begin{pmatrix} \MMhat & \Mbar \\ \Mbar & 1 \end{pmatrix}
\begin{pmatrix} \theta_1 \\ \theta_3 \end{pmatrix}
=
\begin{pmatrix} \YMhat \\ \bar{y}\end{pmatrix}
\end{equation}

which reduces to

\begin{align}
\begin{pmatrix} \theta_1 \\ \theta_3 \end{pmatrix}
&=
\frac{1}{\MMhat - \Mbar^2}
\begin{pmatrix} 1 & -\Mbar \\ -\Mbar & \MMhat \end{pmatrix}
\begin{pmatrix} \YMhat \\ \bar{y}\end{pmatrix}\\
&=
\frac{1}{\MMhat - \Mbar^2}
\begin{pmatrix} \YMhat - \bar{y}\Mbar \\ \MMhat\bar{y} - \YMhat\Mbar \end{pmatrix}
\end{align}

Letting $MM = \MMhat - \Mbar^2$ and $YM = \YMhat - \bar{y}\Mbar$, we have
\begin{equation}
\begin{pmatrix} \theta_1 \\ \theta_3 \end{pmatrix}
=
\begin{pmatrix} YM / MM \\ \bar{y} - \Mbar (YM / MM)\end{pmatrix}
\end{equation}

This means we can rewrite the model $\hat{y} = \theta_1\Mshft + \theta_3$ as

\begin{equation}
\hat{y}_i = \bar{y} + \left(\frac{YM}{MM}\right)(M_i - \Mbar)
\end{equation}

To obtain an expression for the periodogram, $P = 1 - \chi^2 / \chi^2_0$, we first compute $\chi^2$

\begin{align}
\chi^2 &= \sum_i w_i (y_i - \hat{y}_i)^2 \\
       &= \sum_i w_i (y_i^2 - 2y_i\hat{y}_i + \hat{y}_i^2)\\
       &= YY - 2\frac{(YM)^2}{MM} + \frac{(YM)^2}{MM}\\
       &= YY - \frac{(YM)^2}{MM}
\end{align}

Since, $\chi^2_0 = YY$, we have

\begin{equation}
P(\omega) = \frac{(YM)^2}{YY\cdot MM}
\end{equation}

We wish to maximize $P(\omega)$ with respect to the phase shift parameter $\theta_2$,

\begin{align}
\label{eq:theta2cond}
\partial_{\theta_2}P = 0 &= \frac{YM}{YY\cdot MM}\left(2\partial_{\theta_2}(YM) - \frac{YM}{MM}\partial_{\theta_2}(MM)\right)\\
                         &= 2MM\partial_{\theta_2}(YM) - YM\partial_{\theta_2}(MM).
\end{align}

The final expression is the non-linear condition that must be satisfied by the optimal
phase shift parameter $\theta_2$. However, satisfying Equation \ref{eq:theta2cond}
is not \emph{sufficient} to guarantee that $\theta_2$ is optimal. The value of the
periodogram at each $\theta_2$ satisfying Equation \ref{eq:theta2cond} must be computed,
and the globally optimal solution chosen from this set.

We seek a more explicit form for Equation \ref{eq:theta2cond}. We
derive expressions for $MM$ and $YM$, defining

\begin{eqnarray}
CC_{nm} &\equiv& \scov(\cos{n\omega t},\cos{m\omega t})\label{eq:CCdef}\\
CS_{nm} &\equiv& \scov(\cos{n\omega t},\sin{m\omega t})\label{eq:CSdef}\\
SS_{nm} &\equiv& \scov(\sin{n\omega t},\sin{m\omega t})\label{eq:SSdef}\\
YC_n &\equiv& \savg{(y - \bar{y})\cos{n\omega t}}\label{eq:YCdef}\\
YS_n &\equiv& \savg{(y - \bar{y})\sin{n\omega t}}\label{eq:YSdef},
\end{eqnarray}

\noindent all of which can be evaluated efficiently using the NFFT.

The autocovariance of the template values $MM$, is given by

\begin{align}
MM &\equiv \sum_i w_i M_i^2 - \left(\sum_i w_i M_i\right)^2\\
   &= \sum_i w_i \left(\sum_nA_n\cos{\omega n t_i} + B_n\sin{\omega n t_i}\right)^2\\
   &\qquad - \left(\sum_nA_nC_n + B_nS_n\right)^2\\
   &= \sum_{n,m} A_nA_mCC_{nm} \nonumber \\
   &\qquad + (A_nB_mCS_{nm} + B_nA_m(CS^T)_{nm}) \nonumber \\
   &\qquad + B_nB_mSS_{nm}\\
   &= \sum_{n,m} A_nA_mCC_{nm} + 2A_nB_mCS_{nm} \nonumber\\
   &\qquad + B_nB_mSS_{nm},
\end{align}

\noindent using

\begin{equation}
\sum_{n,m}A_nB_mCS_{nm} = \sum_{n,m} A_mB_n CS_{mn}.
\end{equation}

We also derive the products $A_nA_m$, $A_nB_m$, $B_nB_m$:

\begin{align}
\begin{split}
A_nA_m &= \left(\alpha_n\eith^n + \alpha^{*}_n\eith^{-n}\right)\left(\alpha_m\eith^m + \alpha^{*}_m\eith^{-m}\right)\\
       &= \alpha_n\alpha_m \eith^{n+m} + \alpha^{*}_n\alpha_m\eith^{m-n} \\
       &\qquad + \alpha_n\alpha^{*}_m \eith^{n-m} + \alpha^{*}_n\alpha^{*}_m\eith^{-n-m)}
\end{split}\\
\begin{split}
A_nB_m &= -i\left(\alpha_n\eith^n + \alpha^{*}_n\eith^{-n}\right)\left(\alpha_m\eith^m - \alpha^{*}_m\eith^{-m}\right)\\
       &= -i\left\{\alpha_n\alpha_m \eith^{n+m} + \alpha^{*}_n\alpha_m\eith^{m-n} \right.\\
       &\qquad \left. - \alpha_n\alpha^{*}_m \eith^{n-m} - \alpha^{*}_n\alpha^{*}_m\eith^{-n-m)}\right\}
\end{split}\\
\begin{split}
B_nB_m &= -\left(\alpha_n\eith^n - \alpha^{*}_n\eith^{-n}\right)\left(\alpha_m\eith^m - \alpha^{*}_m\eith^{-m}\right)\\
       &= -\left\{\alpha_n\alpha_m \eith^{n+m} - \alpha^{*}_n\alpha_m\eith^{m-n} \right.\\
       &\qquad \left.- \alpha_n\alpha^{*}_m \eith^{n-m} + \alpha^{*}_n\alpha^{*}_m\eith^{-n-m}\right\}
\end{split}
\end{align}

Now we have that

\begin{align}
\begin{split}
MM_{nm} &= A_nA_mCC_{nm} + 2A_nB_mCS_{nm} + B_nB_mSS_{nm}\\
        &= \alpha_n\alpha_m\CCt_{nm}\eith^{n+m} + 2\alpha_n\alpha^{*}_m\CSt_{nm}\eith^{n-m} \\
        &\qquad + \alpha^{*}_n\alpha^{*}_m\SSt_{nm}\eith^{-(n+m)}
\end{split}
\end{align}

where

\begin{align}
\CCt_{nm}  &= (CC_{nm} - SS_{nm}) - i\left(CS + CS^T\right)_{nm} \\
\CSt_{nm}  &= (CC_{nm} + SS_{nm}) + i\left(CS - CS^T\right)_{nm}\\
\SSt_{nm}  &= (CC_{nm} - SS_{nm}) + i\left(CS + CS^T\right)_{nm}
\end{align}

and for $YM$:

\begin{align}
\begin{split}
YM_{k} &= A_kYC_k + B_kYS_k\\
        &= \alpha_kYC_k\eith^k + \alpha_k^{*}YC_k\eith^{-k} \\
        &\qquad - i\left(\alpha_kYS_k\eith^k - \alpha^{*}_kYS_k\eith^{-k}\right)\\
        &= (YC_k - iYS_k)\alpha_k\eith^k + (YC_k + iYS_k)\alpha^{*}_k\eith^{-k}\\
        &= \alpha_k\YCt_k \eith^k + \alpha^{*}_k\YCt_k^{*} \eith^{-k}.
\end{split}
\end{align}

We also define $YM' = \eith^H YM$ and $MM' = \eith^{2H}MM$, both of which are polynomials in $\eith$. Their derivatives
are

\begin{align}
\partial YM' &= H\eith^{H-1} YM + \eith^H \partial YM\\
\partial MM' &= 2H\eith^{2H-1} MM + \eith^{2H}\partial MM.
\end{align}

A new polynomial condition can then be expressed in terms of $MM'$, $YM'$ and their derivatives.

\begin{align}\label{eq:finalpoly}
0 &= 2MM\partial(YM) - YM\partial(MM)\\
  &= \eith^{3H+1}\left(2MM\partial(YM) - YM\partial(MM)\right)\\
  &= 2\eith^{2H}MM\left(\eith^{H+1}\partial(YM)\right) \nonumber \\
  &\qquad - \eith^HYM\left(\eith^{2H+1}\partial(MM)\right)\\
  &= 2MM'\left(\eith\partial(YM') - H(YM')\right) \nonumber \\
  &\qquad - YM'\left(\eith\partial(MM') - 2H(MM')\right)\\
  &= \eith\left(2MM'\partial(YM') - YM'\partial(MM')\right) \nonumber \\
  &\qquad- 2H\left(MM'YM' - YM'MM'\right)\\
  &= 2MM'\partial(YM') - YM'\partial(MM')
\end{align}

The last step assumes that $\eith\ne 0$, which is a valid assumption since $\eith=e^{i\theta_2}$ lies on the
unit circle for all real $\theta_2$.

We solve for the zeros of the polynomial condition defined by Equation \ref{eq:finalpoly}
using the \texttt{numpy.polynomial.polyroots} function, which solves for the eigenvalues
of the polynomial companion matrix.

Solving for the zeros of a polynomial given a set of coefficients is unstable in certain cases, since the
coefficients are represented as floating point numbers with finite precision. Thus, we scale the roots
by their modulus to ensure they lie on the unit circle. Alternatively, we could use iterative schemes
such as Newton's method to improve the estimate of the roots more robustly, however this requires more computational power and
the accuracy of the roots was not a problem for any of the cases the authors have tested.


\subsection{Negative amplitude solutions}
The model $\hat{y}= \theta_1\Mshft + \theta_0$ allows for $\theta_1 < 0$ solutions.
In the original formulation of Lomb-Scargle and in linear
extensions involving multiple harmonics, negative amplitudes translate to
phase differences, since $-\cos{x} = \cos(x - \pi)$ and $-\sin{x} = \sin(x - \pi)$.

However, for non-sinusoidal templates, $\mathbf{M}$, negative amplitudes
do not generally correspond to a phase difference. For example, a
detached eclipsing binary template $\mathbf{M}_{\rm EB}(x)$ cannot be
expressed in terms of a phase-shifted negative eclipsing binary template; i.e.
$\mathbf{M}_{\rm EB} \neq - \mathbf{M}_{\rm EB}(x - \phi)$ for any $\phi\in[0, 2\pi)$.

Negative amplitude solutions found by the fast template periodogram are usually
undesirable, as they may produce false positives for lightcurves that resemble
flipped versions of the desired template, and allowing for $\theta_1 < 0$ solutions
increases the number of effective free parameters of the model, which lowers
the signal to noise, especially for weak signals.

One possible remedy for this problem is to set $P_{\rm FTP}(\omega) = 0$ if the optimal
solution for $\theta_1$ is negative, but this complicates the interpretation of $P_{\rm FTP}$.
Another possible remedy is, for frequencies that have a $\theta_1 < 0$ solution,
to search for the optimal parameters while enforcing that $\theta_1 > 0$,
e.g. via non-linear optimization, but this likely will eliminate the computational
advantage of FTP over existing methods.

Thus, we allow for negative amplitude solutions in the model fit and caution
the user to check that the best fit $\theta_1$ is positive.

\subsection{Extending to multi-band observations}

\subsubsection{Multi-phase model}
\label{sec:multiband}
As shown in \cite{Vanderplas+Ivezic_2015}, the multi-phase periodogram (their
$(N_{\rm base}, N_{\rm band}) = (0, 1)$ periodogram), for any model can
be expressed as a linear combination of single-band periodograms:

\begin{equation}
\label{eq:multibandmultiphase}
P^{(0,1)}(\omega) = \frac{\sum_{k=1}^K\chi^2_{0, k}P_{k}(\omega)}{\sum_{k=1}^K\chi^2_{0,k}}
\end{equation}

\noindent where $K$ denotes the number of bands, $\chi^2_{0,k}$ is the weighted sum of squared
residuals between the data in the $k$-th band and its weighted mean $\savg{y}$, and $P_k(\omega)$ is
the periodogram value of the $k$-th band at the trial frequency $\omega$.

With Equation \ref{eq:multibandmultiphase}, the template periodogram is readily applicable to multi-band
time series, which is crucial for experiments like LSST, SDSS, Pan-STARRS, and other current
and future photometric surveys.

Other multi-band extensions of the template periodogram are provided in Appendix \ref{sec:shphmult}.

\subsection{Computational requirements}\label{sec:compreqs}

For a given number of harmonics $H$, the task of deriving
the polynomial given in Equation \ref{eq:finalpoly} requires $\bigO(H^2)$ computations,
and finding the roots of this polynomial requires $\bigO(H^3)$ computations. The degree of the final
polynomial is $6H - 1$.

When considering $N_f$ trial frequencies, the polynomial computation and root-finding
step scales as $\bigO(H^3N_f)$. The computation of the sums
(Equations \ref{eq:CCdef} -- \ref{eq:YSdef}) scales as $\bigO(HN_f\log HN_f)$.

Therefore, the entire template periodogram scales as

\begin{equation}
\bigO(HN_f \log HN_f + H^3N_f).
\end{equation}

However, an important consideration is that the peak width scales inversely with the number of harmonics $\delta f_{\mathrm{peak}} \propto 1/H$ and so the number of trial frequencies needed to resolve a peak increases linearly with the number of harmonics in the template.

The computational scaling factoring in an extra power of $H$ is therefore

\begin{equation}
\bigO(H^2N_f \log HN_f + H^4N_f).
\end{equation}

\begin{figure}
    \centering
    \includegraphics[width=0.5\textwidth]{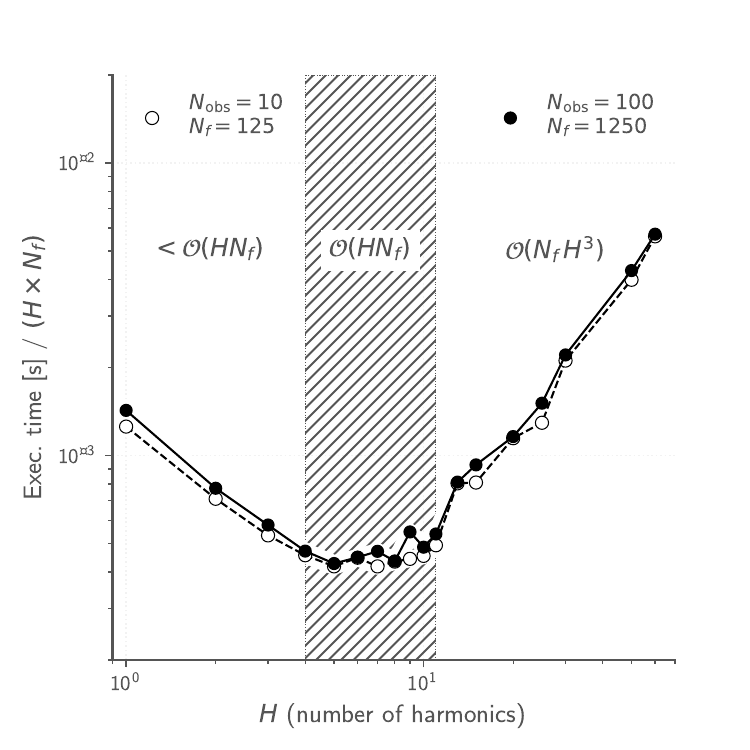}
    \caption{\label{fig:timingnharm} Computation time of FTP scaled by $NH$ for
            different numbers of harmonics. For $H\lesssim 3$, FTP scales
            sublinearly in $H$ (possibly due to a constant overhead per
            trial frequency, independent of $H$). When $3 \lesssim H \lesssim 11$,
            FTP scales approximately linearly in $H$, and when $H \gtrsim 11$
            FTP approaches the $\bigO(H^3)$ scaling limit.}
\end{figure}

For a fixed number of harmonics $H$, the template periodogram scales as
$\bigO(N_f\log N_f)$. However, for a constant number of trial frequencies $N_f$,
the template algorithm scales as $\bigO(H^3)$, and computational resources
alone limit $H$ to reasonably small numbers $H\lesssim15$ (see Figure \ref{fig:timingnharm}).



\section{Implementation}\label{sec:implementation}

\begin{figure*}
    \centering
    \includegraphics[width=\textwidth]{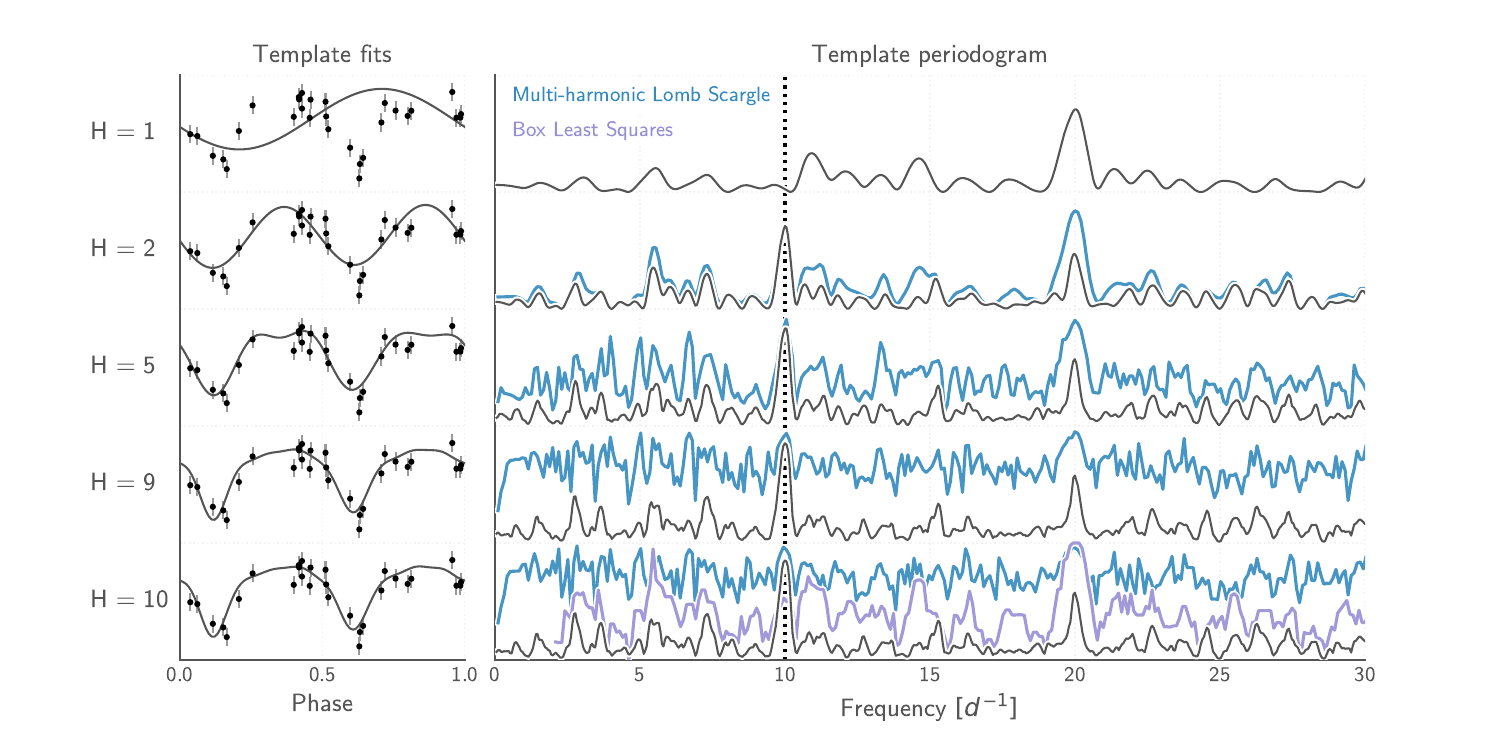}
    \caption{\label{fig:tempsandpdgs} Template periodograms performed on a simulated eclipsing
            binary lightcurve (shown phase-folded in the left-hand plots). The top-most plot
            uses only one harmonic, equivalent to a Lomb-Scargle periodogram. Subsequent plots
            use an increasing number of harmonics, which produces a narrower and higher peak
            height around the correct frequency. For comparison, the multi-harmonic extension
            to Lomb-Scargle is plotted in blue, using the same number of harmonics as the FTP.
            The Box Least-Squares \citep{Kovacs_2002} periodogram is shown in the final plot.}
\end{figure*}

An open-source implementation of the template periodogram in Python is
available.\footnote{\url{https://github.com/PrincetonUniversity/FastTemplatePeriodogram}}
Polynomial algebra is performed using the \texttt{numpy.polynomial} module
\citep{Scipy}. The \texttt{nfft} Python module,
\footnote{\url{https://github.com/jakevdp/nfft}} which provides a Python
implementation of the non-equispaced fast Fourier transform,
is used to compute the necessary sums for a particular time series.

No explicit parallelism is used anywhere in the current implementation,
however certain linear algebra operations in \texttt{Scipy} use OpenMP
via calls to BLAS libraries that have OpenMP enabled.

All timing tests were run on a quad-core 2.6 GHz Intel Core i7 MacBook
Pro laptop (mid-2012 model) with 8GB of 1600 MHz DDR3 memory. The \texttt{Scipy} stack
(version 0.18.1) was compiled with multi-threaded MKL libraries.

\subsection{Comparison with non-linear optimization}

In order to evaluate the accuracy and speed of the template periodogram,
we have included slower alternative solvers within the Python implementation
of the FTP that employ non-linear optimization to find the best fit parameters.

Periodograms computed in Figures \ref{fig:tempsandpdgs}, \ref{fig:corrwgats},
and \ref{fig:corrwhighh} used simulated data. The simulated data has uniformly
random observation times, with Gaussian-random, homoskedastic, uncorrelated
uncertainties. An eclipsing binary template, generated by fitting a well-sampled,
high signal-to-noise eclipsing binary in the HATNet dataset (BD+56 603)
with a 10-harmonic truncated Fourier series.

\subsubsection{Accuracy}

\begin{figure}
    \centering
    \includegraphics[width=0.5\textwidth]{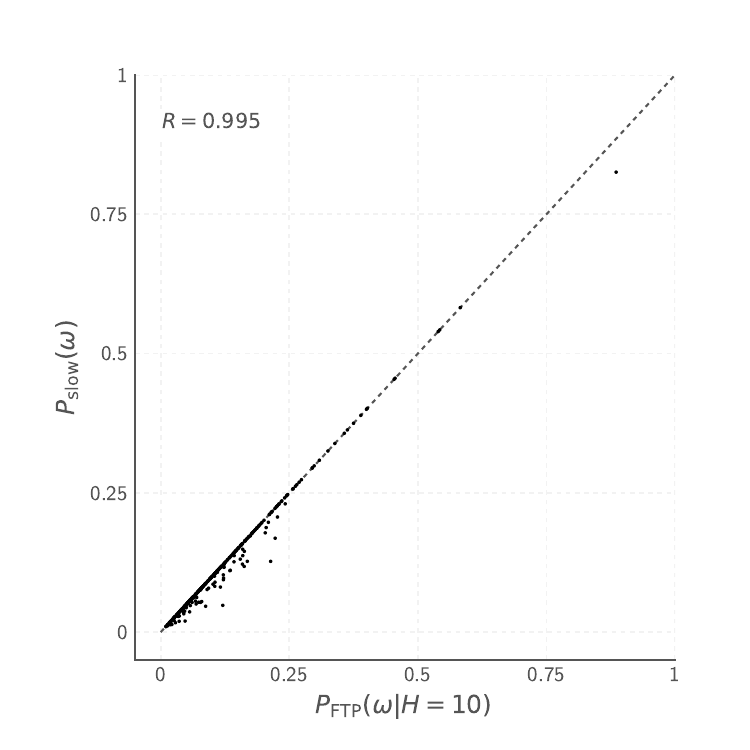}
    \caption{\label{fig:corrwgats} Comparing accuracy between previous methods that rely
            on non-linear optimization at each trial frequency with the fast template periodogram
            described in this paper. Both methods are applied to the same simulated data as shown
            in Figure \ref{fig:tempsandpdgs}.
            The FTP consistently finds more optimal template fits than
            those found with non-linear optimization, which do not guarantee convergence to
            a globally optimal solution. The FTP solves for the optimal
            fit parameters directly, and therefore is able to achieve greater accuracy than template
            fits done via non-linear optimization.}
\end{figure}

\begin{figure*}
    \centering
    \includegraphics[width=\textwidth]{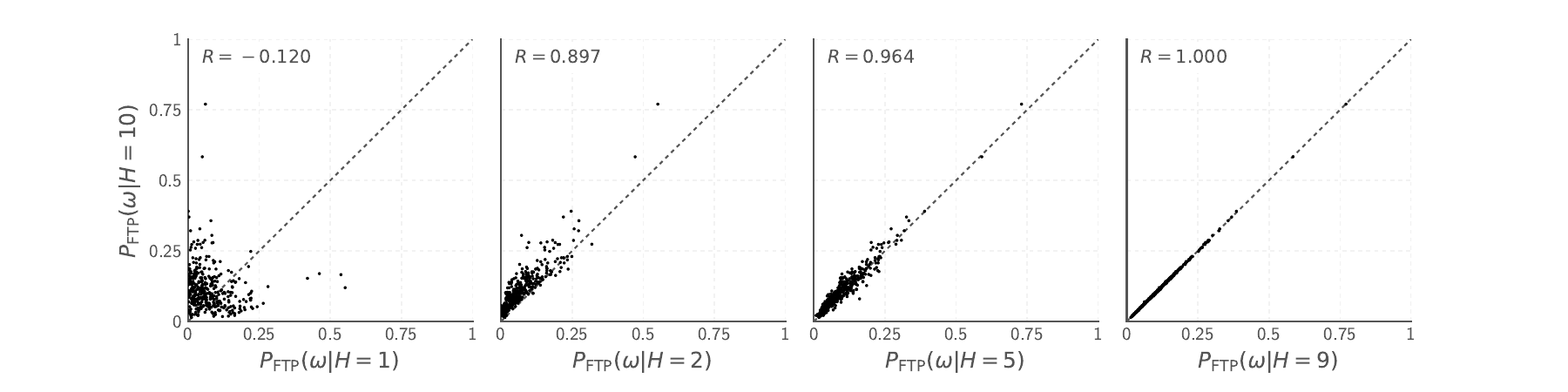}
    \caption{\label{fig:corrwhighh} Comparing the template periodogram calculated with $H=10$ harmonics
            to the template periodogram using a smaller number of harmonics $H < 10$. The template and
            data used to perform the periodogram calculations are the same as those shown in Figure \ref{fig:tempsandpdgs}.}
\end{figure*}

For weak signals or signals folded at the incorrect trial period, there
may be a large number of local $\chi^2$ minima in the parameter space, and thus
non-linear optimization algorithms may have trouble finding the global minimum. The
FTP, on the other hand, solves for the optimal parameters directly, and
thus is able to recover optimal solutions even when the signal is weak
or not present.

Figure \ref{fig:corrwgats} illustrates the accuracy improvement with FTP.
Many solutions found via non-linear optimization are significantly suboptimal
compared to the solutions found by the FTP.

Figure \ref{fig:corrwhighh} compares FTP results obtained using the full template
$(H=10)$ with those obtained using smaller numbers of harmonics. The left-most
plot compares the $H=1$ case (weighted Lomb-Scargle), which, as also demonstrated
in Figure \ref{fig:tempsandpdgs}, illustrates the advantage of the template
periodogram for known, non-sinusoidal signal shapes.

\subsubsection{Computation time}

\begin{figure*}
    \centering
    \includegraphics[width=0.45\textwidth]{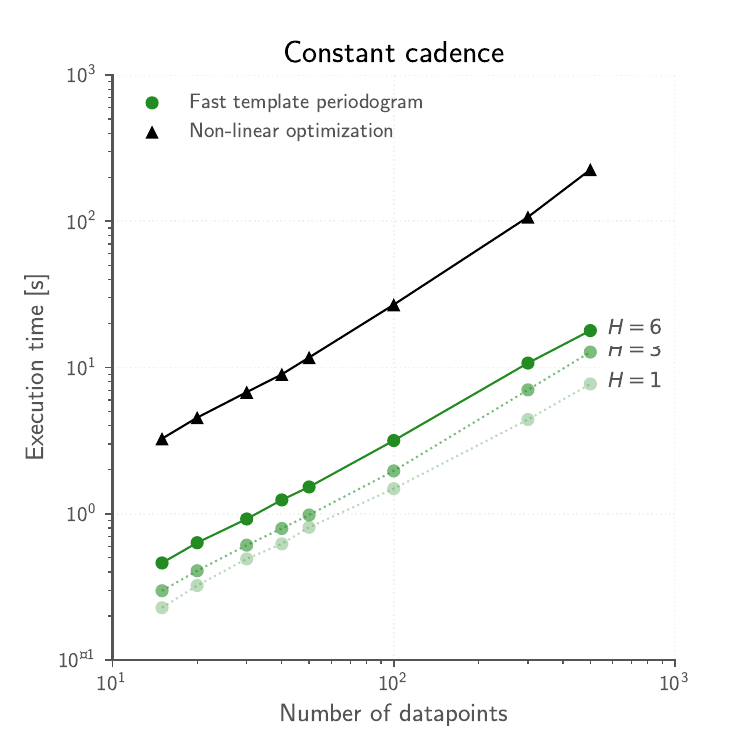}
    \includegraphics[width=0.45\textwidth]{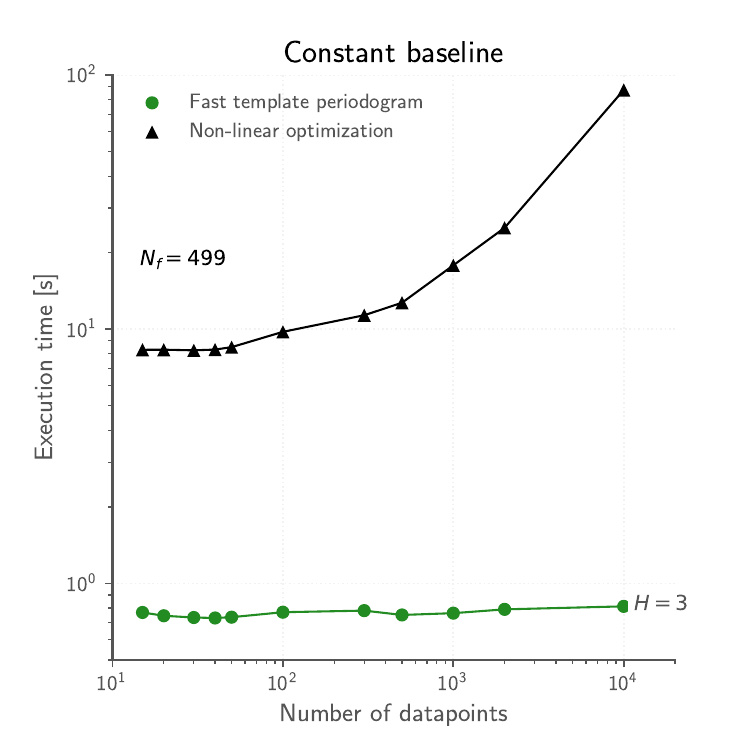}
    \caption{\label{fig:timingndata} Computation time of FTP compared with alternative techniques
             that use non-linear optimization at each trial frequency. \emph{Left}: timing for the
             case when $N_f = ~12 N_{\rm obs}$, i.e. the cadence of the observations is constant.
             \emph{Right}: timing for the case when $N_f$ is fixed, i.e. the baseline of the observations
             are constant. Non-linear optimization techniques scale as $\bigO(N_f N_{\rm obs})$ while
             the FTP scales as $\bigO(HN_f\log HN_f + N_fH^3)$, where $H$ is the number of harmonics
             needed to approximate the template.}

\end{figure*}

FTP scales asymptotically as $\bigO(N_fH\log N_fH)$ with respect to the number of trial frequencies,
$N_f$ and as $\bigO(N_fH^3)$ with respect to the number of harmonics in which the template is expanded, $H$.
For a given resolving power, however, there is an additional factor of $H$ in each of these terms due to the
number of trial frequencies necessary to resolve a periodogram peak being proportional to $H$.
However, for reasonable cases ($N_f \lesssim 10^{120}$ when $H=5$) the computation time is dominated by
computing polynomial coefficients and root finding, both of which scale linearly in $N_f$.

The number of trial frequencies needed for finding astrophysical signals in a typical photometric time series is

\begin{equation}
    N_f = 1.75 \times 10^6 \left(\frac{H}{1}\right)\left(\frac{{\rm baseline}}{10~{\rm yrs}}\right)\left(\frac{\alpha}{5}\right)\left(\frac{15~{\rm mins}}{P_{\rm min}}\right)
\end{equation}

\noindent where $\alpha$ represents the ``oversampling factor,'' $\Delta f_{\rm peak} / \Delta f$, where $\Delta f_{\rm peak} \sim 1/{\rm baseline}$ is
the typical width of a peak in the periodogram and $\Delta f$ is the frequency spacing of the periodogram.

Extrapolating from the timing of a test case (500 observations, 5 harmonics, 15,000 trial frequencies), the summations
account for approximately 5\% of the computation time when $N_f \sim 10^6$. If polynomial computations and
root-finding can be improved to the point where they no longer dominate the computation time, this would provide an
order of magnitude speedup over the current implementation.

Figure \ref{fig:timingndata} compares the timing of the FTP with that of previous methods that employ
non-linear optimization. For the case when $N_f\propto N_{\rm obs}$, FTP achieves a factor of 3 speedup for even
the smallest test case (15 datapoints), while for larger cases ($N\sim10^4$) FTP offers 2-3 orders of magnitude
speed improvement. For the constant baseline case, FTP is a factor of $\sim 2$ faster for the smallest test case
and a factor of $\sim 20$ faster for $N_{\rm obs}\sim 10^4$. Future improvements to the FTP implementation
could further improve speedups by 1-2 orders of magnitude over non-linear optimization.

%

%

\section{Discussion}\label{sec:discussion}

Template fitting is a powerful technique for accurately recovering
the period and amplitude of objects with \emph{a priori} known
lightcurve shapes. It has been used in the literature by, e.g.
\cite{Stringer_et_al_2019,Sesar_etal_2016, Sesar_etal_2010}, to analyze RR Lyrae in the
SDSS, PS1, and DES datasets, where it has been shown to produce purer
samples of RR Lyrae at a given completeness. The computational
cost of current template fitting algorithms, however, limits their
application to larger datasets or with a larger number of templates.

We have presented a novel template fitting algorithm that extends
the Lomb-Scargle periodogram \citep{Lomb_1976,Scargle_1982,Barning_1963,Vanicek_1971}
to handle non-sinusoidal signals that can be expressed in terms of
a truncated Fourier series with a reasonably small number of harmonics
($H\lesssim 10$).

The fast template periodogram (FTP) asymptotically scales as
$\bigO(N_fH^2\log N_fH^2 + N_fH^4)$, while previous template fitting algorithms
such as the one used in the \texttt{gatspy} library \citep{gatspy},
scale as $\bigO(N_fN_{\rm obs})$. However, the FTP effectively
scales as $\bigO(N_fH^4)$, since the time needed to compute polynomial
coefficients and perform zero-finding dominates the computational time
for all practical cases ($N_f \lesssim 10^{120}$).
The $H^4$ scaling effectively restricts templates to those that are
sufficiently smooth to be explained by a small number of Fourier terms.

FTP also improves the accuracy of previous template fitting algorithms,
which rely on non-linear optimization at each trial frequency to minimize
the $\chi^2$ of the template fit. The FTP routinely finds superior fits over
non-linear optimization methods.

An open-source Python implementation of the FTP is available at
GitHub.\footnote{\url{https://github.com/PrincetonUniversity/FastTemplatePeriodogram}}
The current implementation could likely be improved by:

\begin{enumerate}
    \item Improving the speed of the polynomial
          coefficient calculations and the zero-finding steps. This could potentially yield
          a speedup of $\sim1-2$ orders of magnitude over the current implementation.
    \item Exploiting the embarassingly parallel nature of the FTP using GPU's.
\end{enumerate}

For a constant baseline, the current implementation improves existing methods by factors
of a $\sim$few for lightcurves with $\bigO(100)$ observations, and by an order of magnitude
or more for objects with more than 1,000 observations. These improvements, taken at face
value, are not enough to make template fitting feasible on LSST-sized datasets. However,
optimizing the polynomial computations could yield a factor of $\sim 25-100$ speedup over
the current implementation, which would make the FTP 1-3 orders of magnitude faster than
alternative techniques.

\begin{appendix}

\section{Shared-phase multi-band template periodogram}\label{sec:shphmult}

We derive a multi-band extension for the template periodogram for data taken in $K$ filters, with $N_k$ observations in the $k$-th filter. We use the same model as the one described in \cite{Sesar_etal_2016} in order to illustrate the applicability of the template periodogram to more sophisticated scenarios.

The $i$-th observation in the $k$-th filter is denoted $\yk_i$. We wish to fit a periodic, multi-band template $\Mt = (\Mtk[1], \Mtk[2], ..., \Mtk[K]): [0, 2\pi)\rightarrow \mathbb{R}^K$ to all observations. We assume the same model used by \cite{Sesar_etal_2016}, which assumes the relative amplitudes, phase shifts, and offsets are shared across bands:

\begin{equation}
\yhk(t|\theta) = \theta_1\Mtk(\omega t - \theta_2) + \theta_3 + \lk
\end{equation}

where $\lk$ is a fixed relative offset for band $k$. The $\chi^2$ for this model is

\begin{equation}
\chi^2 = \sum_{k=1}^{K}\sum_{i=1}^{N_k}\wk_i \left(\yk_i - \yhk_i\right)^2
\end{equation}

To make things simpler, we can set the $\lk$ values to 0 simply by subtracting them off from our observations; this means we take all $\yk_i \rightarrow \yk_i - \lk$. We have that

\begin{equation}
\chi^2 = \sum_{k=1}^K \Wtk\chi^2_k
\end{equation}

Where $\Wtk \equiv \sum_{i=1}^{N_k}\wk_i$ and $\sum_{k=1}^K\Wtk = 1$. This means that the system of equations reduces to:

\begin{equation}
0 = \frac{\partial\chi^2}{\partial\theta_1} = \sum_{k=1}^K \Wtk\left(\YMh - \theta_1\MMh - \theta_3\Mbk\right)
\end{equation}

for the $\theta_1$ parameter, where $\YMh$, $\MMh$, $\Mbk$ are values from the single band case computed for each band individually, holding $\Wtk = 1$ for each band. That is:

\begin{align}
\YMh &\equiv \frac{1}{\Wtk}\sum_{i=1}^{N_k}\wk_i \yk_i \Mtshftk(\omega t_i)\\
\MMh &\equiv \frac{1}{\Wtk}\sum_{i=1}^{N_k}\wk_i \left(\Mtshftk(\omega t_i)\right)^2\\
\Mbk &\equiv \frac{1}{\Wtk}\sum_{i=1}^{N_k}\wk_i \Mtshftk(\omega t_i)
\end{align}

For the offset $\theta_3$, we have

\begin{equation}
0 = \frac{\partial\chi^2}{\partial\theta_3} = \sum_{k=1}^K \Wtk\left(\yb - \theta_1\Mbk - \theta_3\right)
\end{equation}

Where $\yb$ is the weighted-mean for the $k$-th band ($\yb \equiv (1/\Wtk)\sum_{i=1}^{N_k}\wk_i\yk_i$), again with $\yk_i \rightarrow \yk_i - \lk$.

So if we redefine the quantities $\YMhat$, $\MMhat$, $\Mbar$, and $\bar{y}$ as weighted averages across the bands, i.e. $\YMhat = \sum_{k=1}^K \Wtk \YMh$, etc., the solution for the optimal parameters has the same form:

\begin{align}
\theta_1 &= \left(\frac{YM}{MM}\right)\\
\theta_3 &= \bar{y} - \Mbar\left(\frac{YM}{MM}\right)
\end{align}

which means that the model for the $k$-th band is

\begin{equation}
\yhk = \bar{y} + \left(\frac{YM}{MM}\right)\left(M^{(k)}_i - \Mbar\right)
\end{equation}

The form of the periodogram has the same form as the single-band case:

\begin{align}
\chi^2 &=\sum_{k=1}^{K}\sum_{i=1}^{N_k}\wk_i \left(\yk_i - \yhk_i\right)^2\\
       &=\sum_{k=1}^{K}\sum_{i=1}^{N_k}\wk_i \left((\yk_i - \bar{y}) - \left(\frac{YM}{MM}\right)(M_i^{(k)} - \Mbar)\right)^2\\
       &=\sum_{k=1}^{K}\sum_{i=1}^{N_k}\wk_i \left((\yk_i - \bar{y})^2 - 2\left(\frac{YM}{MM}\right)(M_i^{(k)} - \Mbar)(\yk_i - \bar{y}) + \left(\frac{YM}{MM}\right)^2(M_i^{(k)} - \Mbar)^2\right)^2\\
       &=YY - 2\frac{(YM)^2}{MM} + \frac{(YM)^2}{MM}\\
       &=YY - \frac{(YM)^2}{MM}.
\end{align}

Since there is a single shared offset between the bands (i.e. we assume the mean magnitude is the same in all bands after subtracting $\lk$), the variance for the signal, $YY$, is not $\sum_{k=1}^{K}\sum_{i=1}^{N_k}\wk_i (\yk_i - \yb)^2$ but $\sum_{k=1}^{K}\sum_{i=1}^{N_k}\wk_i (\yk_i - \bar{y})^2$.

Construction of the polynomial for the multi-band case can be performed by first computing the polynomial expression for $\eith^{2H}\MMhat = \sum_{k=1}^{K}\sum_{i=1}^{N_k}\wk_i \left(\eith^{H}\Mk_i\right)^2$ and a separate polynomial expression for $\psi^H\Mbar$, which can then be squared and subtracted from $\psi^{2H}\MMhat$ to find $MM' = \eith^{2H}MM$.

For $YM' = \eith^H YM$, we merely compute $\eith^H\YMhat^{(k)}$ for each band, take the weighted average of the polynomial coefficients for all bands ($\eith^H\YMhat = \sum_k^K \Wtk \eith^H\YMhat^{(k)}$) and subtract the $\eith^H\bar{y}\Mbar$ polynomial to get $YM'$.

After computing the polynomials $MM'$ and $YM'$, the polynomial in Equation \ref{eq:finalpoly} can be computed quickly and the following steps for finding the optimal model parameters is the same as in the single band case.

The computational complexity of this model scales as $\bigO(KN_fH\log N_fH + N_fH^3)$ where $K$ is the number of filters. Since the polynomial zero-finding step is the limiting computation in most real-world applications, the multi-band template periodogram corresponding to the \cite{Sesar_etal_2016} model should not be significantly more computationally intensive than the single-band case.

\section{Data availibility}
The data underlying this article are available in the GitHub repository located at \href{https://github.com/PrincetonUniversity/FastTemplatePeriodogramPaper}{https://github.com/PrincetonUniversity/FastTemplatePeriodogramPaper}.

\end{appendix}

\begin{acknowledgements}
    Joel Hartman and GB acknowledge support from NASA grant NNX17AB61G. JTV is supported by the University of Washington eScience Institute, with funding from the Alfred P. Sloan Foundation, the Gordon and Betty Moore Foundation, and the Washington Research Foundation.
\end{acknowledgements}

\bibliographystyle{apj}
\bibliography{refs}
\end{document}